\documentclass[apsrev,prbrc,twocolumn,showpacs]{revtex4}

\usepackage{graphicx}
\usepackage{dcolumn}
\usepackage{amsmath}

\begin{document}

\title[Short Title]{
Anisotropic Decay Dynamics of Photoexcited Aligned Carbon
Nanotube Bundles}

\author{Y.~Hashimoto,$^{1,2,\dagger}$
 Y.~Murakami,$^{1,3}$
S.~Maruyama,$^{3}$ and J.~Kono$^{1}$}
\thanks{To whom correspondence should be addressed.}
\homepage{http://www.ece.rice.edu/~kono}
\email{kono@rice.edu}

\smallskip

\affiliation{$^{1}$Department of Electrical and Computer
Engineering, Rice University, Houston, Texas
77005, USA\\
$^{2}$Graduate School of Science and Technology and
Venture Business Laboratory, Chiba University,
Chiba 263-8522, Japan\\ $^{3}$Department of Mechanical
Engineering, University of Tokyo 7-3-1 Hongo, Bunkyo-ku, Tokyo
113-8656, Japan}

\date{\today}

\begin{abstract}
We have performed polarization-dependent ultrafast pump-probe
spectroscopy of a film of aligned single-walled carbon nanotube
bundles.
By taking into account imperfect nanotube alignment as well as
anisotropic absorption cross sections, we quantitatively
determined distinctly different photo-bleaching dynamics for
polarizations parallel and perpendicular to the tube axis.  For
perpendicular polarization, we observe a slow (1.0-1.5~ps)
relaxation process, previously unobserved in randomly-oriented
nanotube bundles.  We attribute this slower dynamics to the
excitation and relaxation of surface plasmons in the
radial direction of the nanotube bundles.  \end{abstract}

\maketitle

There is currently much interest in single-walled carbon
nanotubes (SWNTs), both from scientific and technological points
of view.  Their unprecedentedly small diameters, combined with
their unique molecular perfection, provide an ideal
one-dimensional (1-D) playground for the exploration of new
physical phenomena and novel applications.  Their
unique mechanical, electronic, optical, and magnetic properties
are under intensive investigations world
wide~\cite{Oconnell06Book}.

Various types of recent optical experiments,
including ultrafast
spectroscopy~\cite{HagenetAl04APA,OstojicetAl04PRL,HuangetAl04PRL,MaetAl04JCP,WangetAl04PRL,EllingsonetAl05PRB,HippleretAl04PCCP,ReichetAl05PRB,ShengetAl05PRB,ManzonietAl05PRL,ChouetAl05PRB,RussoetAl06PRB,HirorietAl06PRL},
have revealed some basic aspects of 1-D excitons with large
binding energies in individualized semiconducting SWNTs.
However, there has been very limited success in experimentally
elucidating the intrinsically-anisotropic optical properties of
SWNTs, despite the large number of detailed theoretical studies.
This is primarily due to the lack of samples that contain
aligned SWNTs of macroscopic sizes.  Recently, vertically-aligned
SWNT (VA-SWNT) films have been
grown~\cite{MurakamietAl04CPL,MaruyamaetAl05CPL,MurakamietAl05Carbon}
using the alcohol chemical vapor deposition
method~\cite{MaruyamaetAl02CPL,MurakamietAl03CPL}.  Although the
bundled nature of SWNTs in these films preclude photoluminescence
studies, they have allowed researchers to perform quantitative
studies on anisotropic optical absorption cross sections of
SWNTs~\cite{MurakamietAl05PRL}.  Furthermore, an
array of aligned SWNT bundles should be able to support unique
collective plasma oscillations that cannot exist in individual
SWNTs, which add new dimensions to the already rich optical
properties of SWNTs.

Here we present results of polarization-dependent ultrafast
pump-probe studies of a VA-SWNT film. We excited the
first metallic subband and observed a clear difference
in decay dynamics between parallel and perpendicular
excitations.  For polarization perpendicular to the tube axis,
the decay of photoinduced transmission change exhibited a slow
(1.0-1.5~ps) component.
We attribute this slow dynamics to the relaxation of surface
plasmons excited in the perpendicular direction.

The VA-SWNT film used in this study was catalytically
synthesized on both sides of a 0.5-mm-thick optically-polished
fused quartz
substrate~\cite{MurakamietAl04CPL,MaruyamaetAl05CPL,MurakamietAl05Carbon},
using the alcohol chemical vapor deposition
method~\cite{MaruyamaetAl02CPL,MurakamietAl03CPL}.
The characteristics of this type of VA-SWNT films have been
described in great detail in a previous
publication~\cite{MurakamietAl05Carbon}.  The average nanotube
diameter, determined by high-resolution transmission electron
microscopy, was $\sim$1.9~nm, and most of SWNTs in the sample
were bundles whose diameters ranged from 5 to 15~nm. The
thickness of the sample investigated in the present work was
approximately 1~$\mu$m in total.  Linear optical
properties of this sample, including its anisotropic
absorption properties, have been fully characterized
previously~\cite{MurakamietAl05PRL}.

Polarization-dependent pump-probe experiments were performed
using $\sim$150-fs pulses from a mode-locked Ti:Sapphire laser
operating at 80~MHz.  The wavelength of both pump and probe
beams was 800~nm (or 1.55~eV), which corresponded to the
excitation energy of the first metallic subband of
the sample~\cite{MurakamietAl05PRL}.  The pump-probe intensity
ratio was $\sim$10:1, and the signal-to-noise ratio in
differential transmission $\Delta T/T$ was better than
$10^{-5}$ using a Si photodiode with standard lock-in
techniques.  The pump fluence was kept small (<~640~nJ/cm$^2$)
in order to avoid high-density carrier
effects~\cite{OstojicetAl04PRL,HuangetAl04PRL,WangetAl04PRL},
and the signal was proportional to the pump fluence.  All
measurements were performed at room temperature.

\begin{figure}
\includegraphics [scale=0.62] {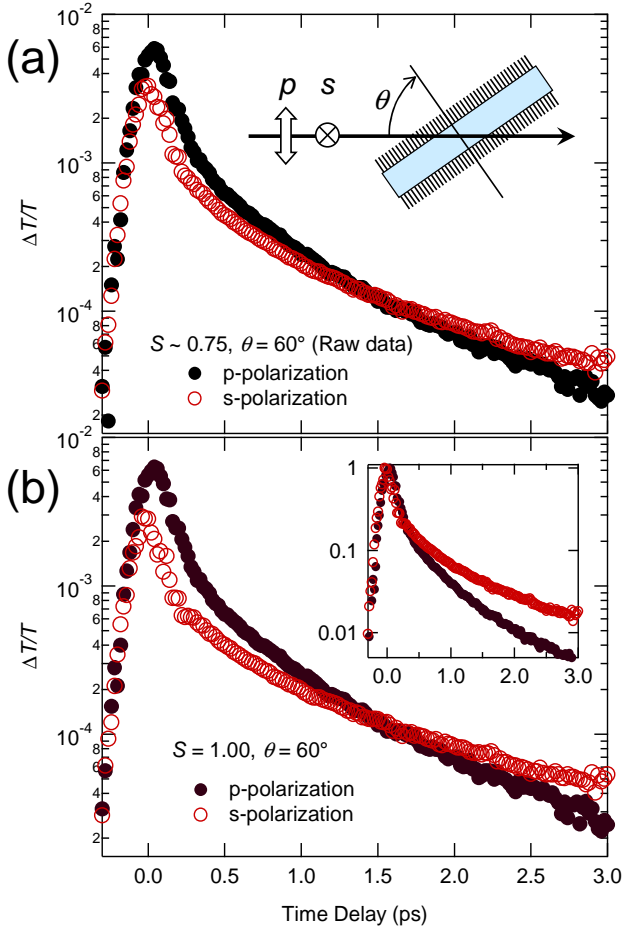}
\caption{Transient changes of transmission of the vertically
aligned SWNT film, whose nematic order parameter $S$ $\sim$ 0.75,
measured with $p$- (filled circle) and $s$-polarizations (open
circles) at photon energy 1.55~eV.  (a) Raw data.  The
inset shows a schematic diagram of the experiment.  (b)
Expected data for the ideal alignment case ($S$ = 1), calculated
from (a) based on linear absorption cross sections.  The inset
shows the same data normalized to the maxima.}
\end{figure}

Figure 1(a) shows raw differential transmission data for
$s$- and $p$-polarizations taken at the incident angle $\theta$ =
60$^\circ$, where $\theta$ is measured from the sample normal,
as schematically shown in the inset of Fig.~1(a).
The inset also indicates the direction  of oscillation of the
electric field of light for the two polarizations.  The pump and
probe polarizations were the same for the data shown in this
figure.  For both $s$- and $p$-polarizations, the sign of
$\Delta T/T$ is positive, i.e., photoinduced ``bleaching'' of
absorption.  The curve for the $s$-polarization is exactly the
same as that taken at $\theta$ = 0$^\circ$ (shown later in the
inset of Fig.~3) when scaled by a factor taking into account the
difference in light path lengths, as expected from the
sample's geometry.

In order to deduce quantitative information on anisotropic
relaxation properties from the data in Fig.~1(a), we need to
take into account the imperfect alignment of nanotubes in the
sample.  The nematic order parameter $S$~\cite{Larson99Book} of
the sample was previously determined to be
0.75~\cite{MurakamietAl05PRL}. In addition, the molar absorption
cross sections (in cm$^2$/mole-C) of perfectly aligned SWNT
bundles for light polarized parallel
and perpendicular to the tube axis, $\sigma_{\parallel}$ and
$\sigma_{\perp}$, were previously determined, and
their ratio
$\sigma_{\perp}$/$\sigma_{\parallel}$ is 0.21 at
1.55~eV~\cite{MurakamietAl05PRL}.
Through a full geometrical consideration, we calculated
the overall optical absorption cross sections of a sample with
order parameter $S$ for $s$- and $p$-polarized light at
incident angle $\theta$, $\epsilon_p(\theta,S)$ and
$\epsilon_s(\theta,S)$, respectively; e.g.,
$\epsilon_p(60^\circ,0.75)$ =
0.647$\cdot\sigma_{\parallel}$ + 0.353$\cdot\sigma_{\perp}$
and $\epsilon_{s}(60^\circ,0.75)$ =
0.082$\cdot\sigma_{\parallel}$ + 0.918$\cdot\sigma_{\perp}$ for
the sample studied, whereas
$\epsilon_{p}(60^\circ,1)$ =
0.75$\cdot\sigma_{\parallel}$ + 0.25$\cdot\sigma_{\perp}$
and $\epsilon_{s}(60^\circ,1)$ = $\sigma_{\perp}$
for perfectly aligned ($S$ = 1) bundles.

The linear transmittance of the sample for $p$-polarized
($s$-polarized) light
at $\theta$ = 60$^\circ$ was 35\% (65\%); i.e.,
65\% (35\%) of incident light is absorbed by the sample at this
wavelength.
This total absorption can be decomposed
into the parallel and perpendicular components;
the amounts of light energy absorbed by the two
components are $I_{p,\parallel}$ = 0.583$I_0$ and
$I_{p,\perp}$ = 0.067$I_0$ ($I_{s,\parallel}$ = 0.104$I_0$ and
$I_{s,\perp}$ = 0.246$I_0$), respectively, where $I_0$ is the
incident light energy (in Joules).
Since the overall temporal shape of the pump-probe signal did not
vary with the pump pulse energy, we are in the linear regime.
Thus, we assume that the observed signal can be decomposed into
parallel and perpendicular components as
\begin{equation}
[\Delta T (t)/T]_i^{S=0.75} = I_{i,\parallel}\cdot \beta_{\parallel}(t)
+ I_{i,\perp}\cdot \beta_{\perp}(t)
\end{equation}
where $i = s,p$ and $\beta_{\parallel}(t)$ and $\beta_{\perp}(t)$
are photo-induced differential transmission per unit
pump energy absorbed by the sample (in J$^{-1}$), corresponding
to the $\sigma_{\parallel}$ and $\sigma_{\perp}$ components of
absorption, respectively.
Figure 1(b) shows transient curves  for the ideal case of $S$ =
1.0 (calculated  using the $\beta_{\parallel}$ and
$\beta_{\perp}$ shown in Fig.~2 below), where the linear
transmittances
for $p$- and $s$-polarizations would become 31\% and 72\%,
respectively. Slight but enhanced contrast between these two
curves compared to the case of Fig.~1(a) is noticeable. The
inset compares the transient curves normalized at their maxima,
showing the existence of anisotropy in the magnitude of their
curves especially in time delays > 1~ps.

\begin{figure}
\includegraphics [scale=0.58] {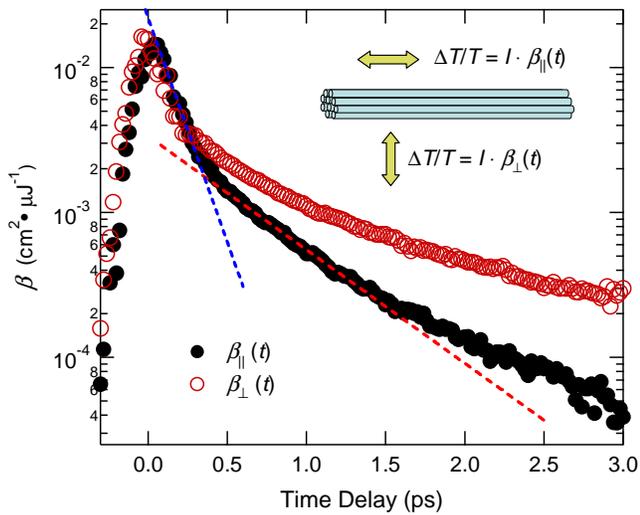}
\caption{Coefficients of the transient transmission caused by
absorption of unit amount of pump energy by SWNTs,
$\beta_{\parallel}(t)$ (filled circles) and $\beta_{\perp}(t)$
(open circles), for excitation polarization parallel and
perpendicular to the SWNT axis, respectively.  See Eq.~(1) for
the definitions. Inset shows a schematic of the corresponding
polarization directions.} \end{figure}

The different dynamics between parallel and perpendicular
excitations can be most clearly seen when
$\beta_{\parallel}(t)$ and $\beta_{\perp}(t)$ are
separately plotted, as shown in Fig.~2. They were obtained
by solving the coupled equations, Eq.~(1) for $i = s,p$,
using the raw $[\Delta T(t) / T]^{S=0.75}_{i}$ data directly from
Fig.~1(a).  Here we clearly see that the perpendicular
contribution has slower relaxation.
It should be noted that $\beta_{\parallel}(t)$ and
$\beta_{\perp}(t)$ have the unit of the inverse of {\em
absorbed} energy density.
Therefore, it is reasonable that they have nearly the same peak
value at $t$ = 0, since the total number of $\pi$
electrons, which are the only electrons excitable at 1.55~eV, is
the same.

The photoinduced transmission signal for excitation parallel to
the SWNT bundles [i.e., $\beta_{\parallel}(t)$] shows faster
decay; 90\% of the decay occurs in the first $\sim$0.5~ps. This
observation is similar to results in prior ultrafast optical
studies of {\em randomly-oriented} bundles of
SWNTs~\cite{HertelMoos00PRL,ChenetAl02APL,IchidaetAl02Physica,LauretetAl03PRL,HanetAl03APL,KorovyankoetAl04PRL,MaedaetAl05PRL,StyersetAl05JPCA}.
Since parallel absorption dominates over perpendicular
absorption in bundled SWNTs ($\sigma_{\parallel}$:$\sigma_{\perp}$ $\approx$  5:1 at
1.55~eV~\cite{MurakamietAl05PRL}), it can be assumed that
pump-probe signals in randomly-oriented bundles are dominated by
carrier dynamics associated with parallel absorption.
There are two dashed lines in Fig.~2, whose slopes correspond to
exponential decay times of 140~fs and 560~fs, respectively.
This type of two-component decay has been invariably observed
in the previous pump-probe studies and variously interpreted.
The fast component can be due to electron-electron
scattering~\cite{HertelMoos00PRL}, intraband
relaxation~\cite{OstojicetAl04PRL,ManzonietAl05PRL}, or
coherent ``artifacts''~\cite{ShengVardeny06PRL}.
The slower component is most likely due to carrier cooling
due to electron-lattice interactions, but the gradual deviation
of the $\beta_{\parallel}$ trace from the 560~fs line means
that the decay process between $\sim$500~fs and 3~ps is not
simple exponential, i.e., the decay becomes slower with time.

\begin{figure}
\includegraphics [scale=0.6] {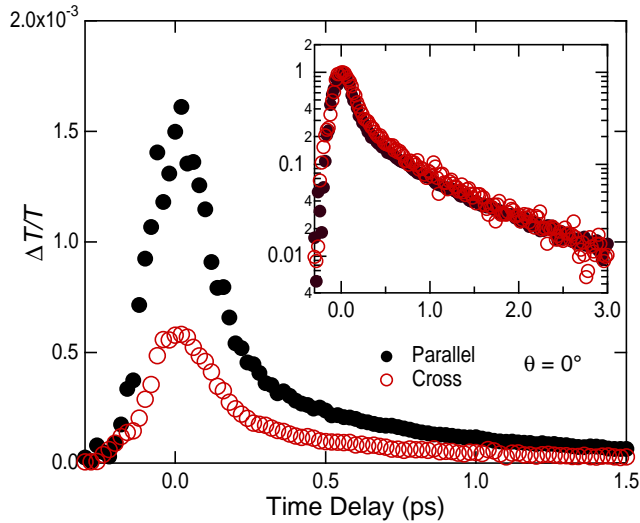}
\caption{Transient changes of transmission measured at normal
incidence measured in ``parallel'' (filled circle) and
``cross'' (open circle) configurations. Inset shows the same
curves normalized to the maxima.} \end{figure}

The slower dynamics observed only in $\beta_{\perp}$, i.e.,
perpendicular excitation, has not been observed
previously, and we regard this as an intrinsic property of the
aligned SWNT bundles associated with {\em charge dynamics in the
radial direction}.  We have further evidence to confirm this
point.  Figure 3 shows pump-probe traces taken at normal
incidence ($\theta$ = 0), i.e., polarization $\perp$ tube axis.
The two traces
were recorded in parallel
and cross configurations, respectively, in regards to the
relative polarizations of the pump and probe beams.  They differ
in magnitude but completely agree with each other in time
dependence, as shown in the inset, where they are normalized to
their respective peak values and plotted in log scale,
exhibiting the characteristic slower component.  The amplitude
ratio $\Delta T_{\parallel} / \Delta T_{\perp}$ is approximately
3.0,
which can be understood through a simple consideration of
excitation of a cylindrical object.  When charge dynamics is
induced by linearly-polarized light oscillating in a direction
perpendicular to the cylinder, the ideal changes in
$T_{\parallel}$ and $T_{\perp}$ due to the depolarization in
the cylinder are given by
\begin{eqnarray}
\Delta T_{\parallel} \propto \int^{2\pi}_{0} \cos^4{\phi}\; d\phi\\
\Delta T_{\perp} \propto \int^{2\pi}_{0} \cos^2{\phi}
\sin^2{\phi} \; d\phi
\end{eqnarray}
where $\phi$ is the azimuthal angle of the cylinder.
The ratio of Eq.~(2) to Eq.~(3) is 3, which is in
excellent agreement with the experimental result.



We propose that the slower dynamics we observed only for light
polarization perpendicular to the tube axis ($\beta_{\perp}$ in
Fig.~2 and both traces in Fig.~3) arise from the excitation and
relaxation of surface plasmons in the SWNT bundles.  Since one
third of SWNTs is metallic, and since one SWNT in a bundle has
usually six neighbors adjacent to it~\cite{ThessetAl96Science},
SWNT bundles can be regarded as conductive wires.  It has been
well established that optical excitation in conductive
nanowires, whose diameter is small enough compared to
the wavelength of incident light, is dominated by the dipolar
surface plasmon excitation that is a collective excitation of
electrons in their radial direction~\cite{KreibigVollmer95Book}.
The surface plasmon mode in sufficiently long nanowires can only
be excited by an electric field perpendicular to their
axes because it is the only direction in which the excited
electrons are spatially confined by the nanowire
surfaces~\cite{SchideretAl01JAP,BarbicetAl02JAP}.
The observed decay in the order of 1.0-1.5~ps can be viewed as
the characteristic time scale of the decay of the excited
surface plasmons in the SWNT bundles.
The reported relaxation times of surface plasmons in metallic
nanoparticles range from sub-ps to a few
ps and are believed to be determined by electron-phonon
interactions~\cite{LinkElSayed99JPCB,VoisinetAl01JPCB,TokizakietAl94APL,HodaketAl98JPCB,InouyeetAl98PRB}.

However, there are at least two notable differences between SWNT
bundles and metallic nanoparticles/nanowires. First, the diameter
distribution of SWNT bundles (5-15~nm in the present
sample~\cite{MurakamietAl05Carbon}) is typically much broader
than that of metallic nanoparticles~\cite{LinkElSayed99JPCB}.
Second, the sharp and strong extinction peaks in the case of
metallic nanowires arise from plasmon {\em resonance}, in which
the denominator of the extinction coefficient derived
from Mie's theory becomes minimum when $\epsilon_1 =
- \epsilon_m$, especially when $|\epsilon_2|$ is
small~\cite{KreibigVollmer95Book}.  Here,
$\epsilon_1$ and $\epsilon_2$ are the real and imaginary parts,
respectively, of the dielectric function of the
nanoparticle, and $\epsilon_m$ is the dielectric function of the
surrounding medium.
Recently, the values of $\epsilon_1$ and $\epsilon_2$ of SWNTs
were reported for both parallel and perpendicular light
polarizations in a wide spectral range~\cite{FaganetAl06PRL},
according to which $\epsilon_1 \sim 0.5$ at 1.55~eV for
perpendicular polarization.  When $\epsilon_m$ = 1 or
larger and 0 <  $\epsilon_1$ < 1, there is a local-field
{\em enhancement} in the cylinder (i.e., the induced
depolarization field has the same direction as the applied
field)~\cite{KreibigVollmer95Book}.  Thus, although the magnitude
should be weaker than the resonant case, we may still excite a
surface plasmon in the perpendicular direction.



In summary, we have studied the anisotropic decay dynamics of
photo-excited aligned SWNT bundles.   By exciting and probing
the first metallic subband, we observed a clear
difference in decay dynamics between parallel and perpendicular
excitations.  When the SWNT bundles are excited by
perpendicularly-polarized light, the decay of photoinduced
transmission change exhibited a slow (1.0-1.5~ps) component that
was absent in the case of excitation by parallel-polarized light.
Based on the fact that the SWNT bundles can be regarded as
conductive nanowires, we attribute the observed slow
component to the relaxation of excited surface plasmons in the
perpendicular direction. In terms of use of the
vertically-aligned bundled SWNT films as, e.g., ultrafast
saturable
absorbers~\cite{ChenetAl02APL,TatsuuraetAl03AM,YamashitaetAl04OL,SakakibaraetAl05JJAP},
the present results indicate that the combination of aligned SWNT
bundles and incident pump light polarized {\em parallel} to the
SWNT axes can provide the fastest {\em turn-off} speed, due to
the absence of the slowly-decaying component.

\smallskip

This work was supported in part by the Robert A.~Welch Foundation
(through Grant No.~C-1509) and the National Science Foundation
(through Grant Nos.~DMR-0134058 and DMR-0325474).

\smallskip

\noindent $^{\dagger}$Present address: Imaging Science and
Engineering Laboratory, Tokyo Institute of Technology, Yokohama,
Kanagawa 226-8503, Japan



\begin{thebibliography}{34}
\expandafter\ifx\csname natexlab\endcsname\relax\def\natexlab#1{#1}\fi
\expandafter\ifx\csname bibnamefont\endcsname\relax
  \def\bibnamefont#1{#1}\fi
\expandafter\ifx\csname bibfnamefont\endcsname\relax
  \def\bibfnamefont#1{#1}\fi
\expandafter\ifx\csname citenamefont\endcsname\relax
  \def\citenamefont#1{#1}\fi
\expandafter\ifx\csname url\endcsname\relax
  \def\url#1{\texttt{#1}}\fi
\expandafter\ifx\csname urlprefix\endcsname\relax\def\urlprefix{URL }\fi
\providecommand{\bibinfo}[2]{#2}
\providecommand{\eprint}[2][]{\url{#2}}

\bibitem[{\citenamefont{O'Connell}(2006)}]{Oconnell06Book}
See, e.g., \bibinfo{editor}{\bibfnamefont{M.~J.}
\bibnamefont{O'Connell}}, ed.,   \emph{\bibinfo{title}{Carbon
Nanotubes: Properties and Applications}}
(\bibinfo{publisher}{CRC Press, Taylor \& Francis Group},
\bibinfo{address}{Boca Raton}, \bibinfo{year}{2006}).

\bibitem{HagenetAl04APA}
A.~Hagen {\it et al}.,
Appl.~Phys.~A {\bf 78}, 1137 (2004); Phys.~Rev.~Lett.~{\bf 95},
197401 (2005).

\bibitem[{\citenamefont{Ostojic et~al.}(2004)\citenamefont{Ostojic, Zaric,
  Kono, Strano, Moore, Hauge, and Smalley}}]{OstojicetAl04PRL}
\bibinfo{author}{\bibfnamefont{G.~N.} \bibnamefont{Ostojic}}
 {\it et al}., \bibinfo{journal}{Phys. Rev. Lett.}
\textbf{\bibinfo{volume}{92}},   \bibinfo{pages}{117402}
(\bibinfo{year}{2004}); {\it ibid}.
\textbf{\bibinfo{volume}{94}},   \bibinfo{pages}{097401}
(\bibinfo{year}{2005}).

\bibitem[{\citenamefont{Huang et~al.}(2004)\citenamefont{Huang, Pedrosa, and
  Krauss}}]{HuangetAl04PRL}
\bibinfo{author}{\bibfnamefont{L.}~\bibnamefont{Huang}},
  \bibinfo{author}{\bibfnamefont{H.~N.} \bibnamefont{Pedrosa}},
  \bibnamefont{and} \bibinfo{author}{\bibfnamefont{T.~D.}
  \bibnamefont{Krauss}}, \bibinfo{journal}{Phys. Rev. Lett.}
  \textbf{\bibinfo{volume}{93}}, \bibinfo{pages}{017403}
  (\bibinfo{year}{2004}); L.~Huang and T.~D.~Krauss,
Phys.~Rev.~Lett.~{\bf 96}, 057407 (2006).

\bibitem[{\citenamefont{Ma et~al.}(2004)\citenamefont{Ma, Stenger, Zimmerman,
  Bachilo, Smalley, Weisman, and Fleming}}]{MaetAl04JCP}
\bibinfo{author}{\bibfnamefont{Y.-Z.} \bibnamefont{Ma}}
{\it et al}.,
  \bibinfo{journal}{J. Chem. Phys.}
  \textbf{\bibinfo{volume}{120}}, \bibinfo{pages}{3368}
(\bibinfo{year}{2004}); Phys.~Rev.~Lett.~{\bf 94}, 157402 (2005);
J.~Phys.~Chem.~B {\bf 109}, 15671 (2005); Phys.~Rev.~B {\bf 74},
085402 (2006).

\bibitem[{\citenamefont{Wang et~al.}(2004)\citenamefont{Wang, Dukovic, Brus,
  and Heinz}}]{WangetAl04PRL}
\bibinfo{author}{\bibfnamefont{F.}~\bibnamefont{Wang}}
{\it et al}.,
\bibinfo{journal}{Phys. Rev. Lett.} \textbf{\bibinfo{volume}{92}},
  \bibinfo{pages}{177401} (\bibinfo{year}{2004});
  Phys.~Rev.~B {\bf 70}, 241403(R) (2004).

\bibitem{EllingsonetAl05PRB}
R.~J.~Ellingson {\it et al}., Phys.~Rev.~B {\bf 71}, 115444
(2005).

\bibitem{HippleretAl04PCCP}
H.~Hippler {\it et al}., Phys.~Chem.~Chem.~Phys.~{\bf 6}, 2387
(2004).

\bibitem{ReichetAl05PRB}
S.~Reich {\it et al}., Phys.~Rev.~B {\bf 71}, 033402
(2005).

\bibitem{ShengetAl05PRB}
C.-X. Sheng {\it et~al}., Phys.~Rev.~B {\bf 71}, 125427 (2005).

\bibitem{ManzonietAl05PRL}
C.~Manzoni {\it et al}., Phys.~Rev.~Lett.~{\bf 94}, 207401
(2005).

\bibitem{ChouetAl05PRB}
S.~G.~Chou {\it et al}., Phys.~Rev.~B {\bf 72}, 195415 (2005).

\bibitem{RussoetAl06PRB}
R.~M.~Russo {\it et al}., Phys.~Rev.~B {\bf 74}, 041405(R)
(2006).

\bibitem{HirorietAl06PRL}
H.~Hirori {\it et~al}., Phys.~Rev.~Lett.~{\bf 97}, 257401 (2006).

\bibitem[{\citenamefont{Murakami et~al.}(2004)\citenamefont{Murakami, Chiashi,
  Miyauchi, Hu, Ogura, Okubo, and Maruyama}}]{MurakamietAl04CPL}
\bibinfo{author}{\bibfnamefont{Y.}~\bibnamefont{Murakami}}
{\it et al}.,
  \bibinfo{journal}{Chem. Phys. Lett.} \textbf{\bibinfo{volume}{385}},
  \bibinfo{pages}{298} (\bibinfo{year}{2004}).

\bibitem[{\citenamefont{Maruyama et~al.}(2005)\citenamefont{Maruyama,
  Einarsson, Murakami, and Edamura}}]{MaruyamaetAl05CPL}
\bibinfo{author}{\bibfnamefont{S.}~\bibnamefont{Maruyama}}
{\it et al}.,
  \bibinfo{journal}{Chem. Phys. Lett.} \textbf{\bibinfo{volume}{403}},
  \bibinfo{pages}{320} (\bibinfo{year}{2005}).

\bibitem[{\citenamefont{Murakami
  et~al.}(2005{\natexlab{a}})\citenamefont{Murakami, Einarsson, Edamura, and
  Maruyama}}]{MurakamietAl05Carbon}
\bibinfo{author}{\bibfnamefont{Y.}~\bibnamefont{Murakami}}
{\it et al}.,
  \bibinfo{journal}{Carbon} \textbf{\bibinfo{volume}{43}},
  \bibinfo{pages}{2664} (\bibinfo{year}{2005}{\natexlab{a}}).

\bibitem[{\citenamefont{Maruyama et~al.}(2002)\citenamefont{Maruyama, Kojima,
  Miyauchi, Chiashi, and Kohno}}]{MaruyamaetAl02CPL}
\bibinfo{author}{\bibfnamefont{S.}~\bibnamefont{Maruyama}}
{\it et al}.,
  \bibinfo{journal}{Chem. Phys. Lett.} \textbf{\bibinfo{volume}{360}},
  \bibinfo{pages}{229} (\bibinfo{year}{2002}).

\bibitem[{\citenamefont{Murakami et~al.}(2003)\citenamefont{Murakami, Miyauchi,
  Chiashi, and Maruyama}}]{MurakamietAl03CPL}
\bibinfo{author}{\bibfnamefont{Y.}~\bibnamefont{Murakami}}
{\it et al}.,
  \bibinfo{journal}{Chem. Phys. Lett.} \textbf{\bibinfo{volume}{374}},
  \bibinfo{pages}{53} (\bibinfo{year}{2003}).

\bibitem[{\citenamefont{Murakami
  et~al.}(2005{\natexlab{b}})\citenamefont{Murakami, Einarsson, Edamura, and
  Maruyama}}]{MurakamietAl05PRL}
\bibinfo{author}{\bibfnamefont{Y.}~\bibnamefont{Murakami}}
{\it et al}.,
  \bibinfo{journal}{Phys. Rev. Lett.} \textbf{\bibinfo{volume}{94}},
  \bibinfo{pages}{087402} (\bibinfo{year}{2005}{\natexlab{b}}).


\bibitem[{\citenamefont{Larson}(1999)}]{Larson99Book}
\bibinfo{author}{\bibfnamefont{R.~G.} \bibnamefont{Larson}},
  \emph{\bibinfo{title}{The Structure and Rheology of Complex Fluids}}
  (\bibinfo{publisher}{Oxford University Press}, \bibinfo{address}{New York},
  \bibinfo{year}{1999}).

\bibitem[{\citenamefont{Hertel and Moos}(2000)}]{HertelMoos00PRL}
\bibinfo{author}{\bibfnamefont{T.}~\bibnamefont{Hertel}} \bibnamefont{and}
  \bibinfo{author}{\bibfnamefont{G.}~\bibnamefont{Moos}},
  \bibinfo{journal}{Phys. Rev. Lett.} \textbf{\bibinfo{volume}{84}},
  \bibinfo{pages}{5002} (\bibinfo{year}{2000});
Chem.~Phys.~Lett.~{\bf 320}, 359 (2000); T.~Hertel, R. Fasel, and
G. Moos, Appl.~Phys.~A {\bf 75}, 449 (2002).

\bibitem[{\citenamefont{Chen et~al.}(2002)\citenamefont{Chen, Raravikar,
  Schadler, Ajayan, Zhao, Lu, Wang, and Zhang}}]{ChenetAl02APL}
\bibinfo{author}{\bibfnamefont{Y.-C.} \bibnamefont{Chen}}
{\it et al}.,
  \bibinfo{journal}{Appl. Phys. Lett.} \textbf{\bibinfo{volume}{81}},
  \bibinfo{pages}{975} (\bibinfo{year}{2002}).

\bibitem{IchidaetAl02Physica}
M.~Ichida {\it et~al}., Physica B {\bf 323}, 237 (2002).

\bibitem[{\citenamefont{Lauret et~al.}(2003)\citenamefont{Lauret, Voisin,
  Cassabois, Delalande, Roussignol, Jost, and Capes}}]{LauretetAl03PRL}
\bibinfo{author}{\bibfnamefont{J.~S.} \bibnamefont{Lauret}}
{\it et al}.,
  \bibinfo{journal}{Phys. Rev. Lett.} \textbf{\bibinfo{volume}{90}},
  \bibinfo{pages}{057404} (\bibinfo{year}{2003}); Phys.~Rev.~B
{\bf 72}, 113413 (2005).

\bibitem{HanetAl03APL}
H.~Han {\it et~al}., Appl.~Phys.~Lett.~{\bf 82}, 1458 (2003).

\bibitem[{\citenamefont{Korovyanko et~al.}(2004)\citenamefont{Korovyanko,
  Sheng, Vardeny, Dalton, and Baughman}}]{KorovyankoetAl04PRL}
\bibinfo{author}{\bibfnamefont{O.~J.} \bibnamefont{Korovyanko}}
{\it et al}.,
 \bibinfo{journal}{Phys. Rev. Lett.}
  \textbf{\bibinfo{volume}{92}}, \bibinfo{pages}{017403}
  (\bibinfo{year}{2004}).

\bibitem{MaedaetAl05PRL}
A.~Maeda {\it et al}., Phys.~Rev.~Lett.~{\bf 94}, 047404 (2005);
J.~Phys.~Soc.~Jpn.~{\bf 75}, 043709 (2006).

\bibitem{StyersetAl05JPCA}
D.~J.~Styers-Barnett {\it et~al}., J.~Phys.~Chem.~A {\bf 109},
289 (2005).

\bibitem{ShengVardeny06PRL}
C.-X.~Sheng and Z.~V.~Vardeny, Phys.~Rev.~Lett.~{\bf 96}, 019705
(2006); H.~Okamoto {\it et al}., {\it ibid}.~{\bf 96}, 019706
(2006).

\bibitem[{\citenamefont{Thess et~al.}(1996)\citenamefont{Thess, Lee, Dai,
  Petit, Robert, Xu, Lee, Kim, Rinzler, Colbert et~al.}}]{ThessetAl96Science}
\bibinfo{author}{\bibfnamefont{A.}~\bibnamefont{Thess}}
{\it et al}., \bibinfo{journal}{Science}
  \textbf{\bibinfo{volume}{273}}, \bibinfo{pages}{483} (\bibinfo{year}{1996}).

\bibitem[{\citenamefont{Kreibig and Vollmer}(1995)}]{KreibigVollmer95Book}
\bibinfo{author}{\bibfnamefont{U.}~\bibnamefont{Kreibig}} \bibnamefont{and}
  \bibinfo{author}{\bibfnamefont{M.}~\bibnamefont{Vollmer}},
  \emph{\bibinfo{title}{Optical Properties of Metal Clusters}}
  (\bibinfo{publisher}{Springer}, \bibinfo{address}{Berlin},
  \bibinfo{year}{1995}).

\bibitem{SchideretAl01JAP}
G.~Schider {\it et~al}., J.~Appl.Phys.~{\bf 90}, 3825 (2001).

\bibitem{BarbicetAl02JAP}
M.~Barbic {\it et~al}., J.~Appl.Phys.~{\bf 91}, 9341 (2002).

\bibitem[{\citenamefont{Link and El-Sayed}(1999)}]{LinkElSayed99JPCB}
\bibinfo{author}{\bibfnamefont{S.}~\bibnamefont{Link}} \bibnamefont{and}
  \bibinfo{author}{\bibfnamefont{M.~A.} \bibnamefont{El-Sayed}},
  \bibinfo{journal}{J. Phys. Chem. B} \textbf{\bibinfo{volume}{103}},
  \bibinfo{pages}{8410} (\bibinfo{year}{1999}).

\bibitem[{\citenamefont{Voisin et~al.}(2001)\citenamefont{Voisin, Fatti,
  Christofilos, and Vallee}}]{VoisinetAl01JPCB}
\bibinfo{author}{\bibfnamefont{C.}~\bibnamefont{Voisin}}
{\it et al}.,
  \bibinfo{journal}{J. Phys. Chem. B} \textbf{\bibinfo{volume}{105}},
  \bibinfo{pages}{2264} (\bibinfo{year}{2001}).

\bibitem[{\citenamefont{Tokizaki et~al.}(1994)\citenamefont{Tokizaki, Nakamura,
  Kaneko, Uchida, Omi, Tanji, and Asahara}}]{TokizakietAl94APL}
\bibinfo{author}{\bibfnamefont{T.}~\bibnamefont{Tokizaki}}
{\it et al}.,
  \bibinfo{journal}{Appl. Phys. Lett.} \textbf{\bibinfo{volume}{65}},
  \bibinfo{pages}{941} (\bibinfo{year}{1994}).

\bibitem[{\citenamefont{Hodak et~al.}(1998)\citenamefont{Hodak, Martini, and
  Hartland}}]{HodaketAl98JPCB}
\bibinfo{author}{\bibfnamefont{J.~H.} \bibnamefont{Hodak}},
  \bibinfo{author}{\bibfnamefont{I.}~\bibnamefont{Martini}}, \bibnamefont{and}
  \bibinfo{author}{\bibfnamefont{G.~V.} \bibnamefont{Hartland}},
  \bibinfo{journal}{J. Phys. Chem. B} \textbf{\bibinfo{volume}{102}},
  \bibinfo{pages}{6958} (\bibinfo{year}{1998}).

\bibitem[{\citenamefont{Inouye et~al.}(1998)\citenamefont{Inouye, Tanaka,
  Tanahashi, and Hirao}}]{InouyeetAl98PRB}
\bibinfo{author}{\bibfnamefont{H.}~\bibnamefont{Inouye}}
{\it et al}.,
  \bibinfo{journal}{Phys. Rev. B} \textbf{\bibinfo{volume}{57}},
  \bibinfo{pages}{11334} (\bibinfo{year}{1998}).


\bibitem{FaganetAl06PRL}
J.~A.~Fagan {\it et al}., unpublished.

\bibitem{TatsuuraetAl03AM}
S.~Tatsuura {\it et~al}., Adv.~Mater.~{\bf 15}, 534 (2003).

\bibitem[{\citenamefont{Yamashita et~al.}(2004)\citenamefont{Yamashita, Inoue,
  Maruyama, Murakami, Yaguchi, Jablonski, and Set}}]{YamashitaetAl04OL}
\bibinfo{author}{\bibfnamefont{S.}~\bibnamefont{Yamashita}}
{\it et al}.,
  \bibinfo{journal}{Optics Lett.} \textbf{\bibinfo{volume}{29}},
  \bibinfo{pages}{1581} (\bibinfo{year}{2004}).

\bibitem[{\citenamefont{Sakakibara et~al.}(2005)\citenamefont{Sakakibara,
  Rozhin, Kataura, Achiba, and Tokumoto}}]{SakakibaraetAl05JJAP}
\bibinfo{author}{\bibfnamefont{Y.}~\bibnamefont{Sakakibara}}
{\it et al}.,
  \bibinfo{journal}{Jpn. J. Appl. Phys.} \textbf{\bibinfo{volume}{44}},
  \bibinfo{pages}{1621} (\bibinfo{year}{2005}).



\end{thebibliography}

\end{document}